\documentclass[useAMS,usenatbib]{mn2e}
\usepackage{epsfig}

\def\HI {H\kern0.1em{\sc i}} 
\def\radm {rad m$^{-2}$} 
\def\etal   {{\rm et~al.\ }}

\title[VLA Polarimetry Observations of PKS\,2322$-$123]{VLA Polarimetry Observations of PKS\,2322$-$123; \\Estimating Magnetic Fields in the Abell 2597 Cluster}
\author[L.~K. Pollack, G.~B. Taylor, \& S.~W. Allen]{L.~K. Pollack,$^{1,2}$\thanks{E-mail: lindsey@astro.ucsc.edu (LKP); gtaylor@nrao.edu (GBT); swa@ast.cam.ac.uk (SWA)} G.~B. Taylor,$^{2,3}$\footnotemark[1] and S.~W. Allen$^{4}$\footnotemark[1]\\
$^{1}$Department of Astronomy \& Astrophysics, University of California at Santa Cruz, Santa Cruz, CA 95064, USA\\
$^{2}$National Radio Astronomy Observatory, Socorro, NM 87801, USA\\
$^{3}$Kavli Institute of Particle Astrophysics and Cosmology, Menlo Park, CA 94025, USA\\
$^{4}$Institute of Astronomy, Madingley Road, Cambridge CB3 0HA, UK}
\begin{document}

\date{Submitted 2004 October 12}
\pagerange{\pageref{firstpage}--\pageref{lastpage}} \pubyear{2004}
\maketitle
\label{firstpage}

\begin{abstract}
We present 5, 8, and 15 GHz total intensity and polarimetric
observations of the radio source PKS\,2322$-$123 taken with the Very
Large Array (VLA).  This small (11 kpc) source is located at the
center of the cooling-core cluster Abell 2597.  The inner X-ray
structure, the radio morphology, and the steep spectral index
($\alpha=-1.8$) in the lobes, all suggest that the radio emission is
confined by the ambient X-ray gas.  We detect a small region of
polarized flux in the southern lobe and are able to calculate a
Faraday rotation measure (RM) of 3620 \radm\ over this region.  Based
on these observations and Chandra X-ray data we suggest
that the southern lobe has been deflected from its original
southwestern orientation to the south and into our line-of-sight.
Using the observed RMs and our calculated electron density profiles,
and assuming both a uniform and tangled magnetic field topology, we
estimate a lower limit of the line-of-sight cluster magnetic field,
$B_{\|}=2.1 \ \mu$G.
\end{abstract}


\section{Introduction}
Strong radio galaxies embedded in galaxy clusters have been used
successfully to probe the intracluster medium (ICM).  Taylor \etal
(1994) defined a sample of 14 clusters found in the flux-limited,
all-sky X-ray sample devised by Edge \etal (1992), and determined that
clusters with nonzero estimated mass flow rates and embedded radio
galaxies with $S_{5000}\ge 100$ mJy commonly display significant
rotation measures (RM), often in excess of 800 \radm, with the
magnitude of the RM proportional to the estimated cooling flow rate.
The observed RMs are thought to be caused by Faraday rotation through
a magnetized cluster gas, and detailed studies of these high RM
sources suggest that minimum cluster magnetic fields of 5 -- $10\mu$G
are common in cooling core clusters \cite{carilli}.

A\,2597 is one of the last in the sample of 14 clusters to be
analyzed.  Edge \etal (1992) classified A\,2597 as a cooling core
cluster, and numerous other observations have confirmed nonzero
cooling flow rates.  Most notably, the analysis of XMM-Newton and
Chandra X-ray data by Morris \& Fabian (2004) inferred a mass cooling
rate of $\sim 50 M_{\odot}$ yr$^{-1}$ across the central regions of
the cluster.  As a cooling core cluster, A\,2597 is expected to have
high-density ambient gas that disturbs the radio jets of embedded
sources \cite{loken}.  Chandra observations by McNamara \etal (2001)
confirm the complex X-ray gas environment, and the detection of
extended broad-line \HI\ absorption toward the embedded radio source
PKS\,2322$-$123 by O'Dea \etal\ (1994) can be attributed to a
population of clouds condensing out of the hot gas, though the mass
rate is only a small fraction of the inferred cooling flow rates.
Very Long Baseline Array (VLBA) observations of PKS\,2322$-$123 made
by Taylor \etal (1999) show a symmetric parsec-scale jet structure.
These observations along with previous radio images by Sarazin \etal
(1995) that reveal a kiloparsec-scale, asymmetric, steep-spectrum jet
morphology support the idea that the radio galaxy is confined and
distorted in the high pressure cluster environment.

Here we discuss our multi-frequency polarimetric VLA observations of the
embedded radio source PKS\,2322$-$123 (z=0.083).  We use the polarized
flux associated with synchrotron radiation from the radio jets to
calculate RMs, and then estimate cluster magnetic fields assuming two
different field topologies.  In \S~\ref{RadioObs} we describe our
radio observations, in \S~\ref{XrayObs} we calculate an electron
density profile based on X-ray observations, in \S~\ref{Results} we
present our results, and in \S~\ref{discussion} we make magnetic field
estimates, suggest a source orientation, and compare our observations
with the literature.  Throughout this paper we assume $H_{\circ}=71$
km s$^{-1}$ Mpc$^{-1}$, $\Omega_{M}=0.270$, and $\Omega_{\rm
vac}=0.730$ so that 1\arcsec=1.542 kpc at z=0.083.

\section{Radio Observations and Data Reduction}\label{RadioObs}

\begin{table}
\caption{Observational Parameters for PKS\,2322$-$123}
\begin{tabular}{l r r r r r r r r}
\hline
Date & Frequency & Bandwidth & Config. & Duration \\
& (MHz) & (MHz) &  & (hours) \\
\hline
\noalign{\vskip2pt}
Jul 1991 & 8415/8465   & 50 & A & 0.11 \\
Aug 1991 & 8415/8465   & 50 & A & 0.11 \\
Jul 1994 & 7815/8165   & 50 & B & 0.69 \\
Jul 1994 & 8515/8885   & 50 & B & 0.61 \\
Jul 1994 & 14785/15185 & 50 & B & 1.39 \\
Dec 1996 & 4985        & 50 & A & 1.18 \\
\hline
\end{tabular}
\end{table}

Very Large Array observations of the radio source PKS\,2322$-$123 were
made at 4.985, 7.815, 8.165, 8.415, 8.465, 8.515, 8.885, 14.785 and
15.185 GHz.  The total time on source is 1.18 hours at 5 GHz, 1.52
hours at 8 GHz, and 1.39 hours at 15 GHz.  Observations in the 5 GHz
band were carried out with the phased VLA.  Details regarding these
observations are given in Table~1.  During the July 1994 observations
the source 3C\,48 was used as the primary flux density calibrator and
3C\,138 was used as an absolute reference for the electric vector
polarization angle (EVPA), $\chi$.  Phase calibration was derived from
the nearby compact source 2345$-$167 with a time between calibrators
of $\sim$40 minutes.  The source 0420+417 was used to obtain
instrumental calibration of polarization leakage terms, and was
observed over a wide range of parallactic angle.  During the December
1996 observations and the July and August 1991 observations, the
source 3C\,286 was used both as the primary flux density calibrator
and as the EVPA calibrator.  Phase calibration was derived from nearby
compact sources.  No polarized intensity images were used from the
1991 and 1996 observations.

The data were reduced in AIPS (Astronomical Image Processing System)
following the standard procedures, and stokes $I$, $Q$ and $U$ images
were produced using the AIPS task IMAGR.  We combined multiple
frequencies within each observing band to produce Stokes $I$ images
with increased sensitivity and better ($u,v$) coverage.  At 8 GHz, the
combined data includes observations with the VLA in both the A and B
configurations.  Where frequencies have been combined, we assume the
resultant frequency to be the average of the included frequencies.
The spectral index images were made using combined frequency images
with matched resolutions.  We define the spectral index, $\alpha$, as
$S_{\nu} \propto \nu^{\alpha}$.

The Faraday rotation measure image was created from the four
frequencies spread out over the 8 GHz observing band in the July 1994
data set.  Data at 5 and 15 GHz were not included in the RM map
because no polarized flux was detected at these frequencies.  For the
four frequencies used in the RM map, polarized intensity, $P$, images
and polarization angle, $\chi$, images were derived from $Q$ and $U$
images.  A pixel by pixel least squares fit of $\chi$ versus
$\lambda^{2}$ gives the RM value and an RM error estimate.  Pixels in
the RM map were flagged if they have an error in $\chi$ greater than
20\degr for any of the four frequencies used.

During data reduction we found a 0.73 mJy compact source, most likely
a background AGN, located at a right ascension of $23^{\rmn{h}}
25^{\rmn{m}} 20\fs94$ and declination of $-12\degr 07\arcmin 05\farcs
52$ (J2000).  In order to reduce the sidelobes from this source, we
made a large image of PKS\,2322$-$123 and cleaned the 0.73 mJy source.

\begin{table}
\caption{Source Properties}
\begin{tabular}{l c c}
\hline
Property & PKS\,2322$-$123 \\
\hline
\noalign{\vskip2pt}
core RA (J2000) & $23^{\rmn{h}} 25^{\rmn{m}} 19\fs7281$ \\
\phantom{core}Dec. (J2000) &  $-12\degr 07\arcmin 27\farcs 230$ \\
\phantom{core}Gal. lat. \phantom{   }($b$) & $-$64.85\degr  \\
\phantom{core}Gal. long. ($l$) & 65.35\degr \\
radial velocity$^{a}$ & 24880 $\pm$ 12 km s$^{-1}$ \\
distance from cluster center &  0.0 Mpc \\
luminosity distance & 373.0 Mpc \\
angular size & 7\arcsec \\
physical size & 10.8 kpc \\
flux density (5 GHz) & 365 mJy \\
power (5 GHz) & 6.08 $\times$ 10$^{24}$ W Hz$^{-1}$ \\
$<$RM$>$ & 3620 \radm \\
$\sigma_{\rm RM}$ & 1080 \radm \\
\hline
\label{SourceProperties}
\end{tabular}

\medskip
$^{a}$Radial velocity is determined from \HI\ absorption observed
toward the nucleus of PKS\,2322$-$123 \cite{tay99}.
\end{table}

\section{X-ray Observations}\label{XrayObs}

McNamara \etal (2001) report X-ray observations of the galaxy cluster
A\,2597 made in July 2000 using the Advanced CCD Imaging Spectrometer 
detector on Chandra. Here, we present a re-analysis of those data
using updated software and calibration products.

The standard level-1 event list produced by the Chandra pipeline
processing was reprocessed using the $CIAO$ (version 3.0.2) software
package, including the latest gain maps and calibration products. Bad
pixels were removed and standard grade selections applied.
Time-dependent gain corrections were applied using A. Vikhlinin's {\it
apply\_gain} routine. The data were cleaned from periods of anomalously
high background using the same author's {\it lc\_clean} script and the
recommended energy ranges and bin sizes for the back-illuminated (S3)
detector. The net exposure time was 6.9ks.

The data have been analysed using the methods described by Allen \etal
(2004). In brief, concentric annular spectra, centred on the radio
source, were extracted from the cleaned events list. Background
spectra were extracted from blank-field background data sets.
Separate photon-weighted response matrices and effective area files
were constructed for each region using the calibration files
appropriate for the time of observation.  The data were analysed using
an enhanced version of the Cambridge image deprojection code and XSPEC
(version 11.3: Arnaud 1996). We have used the MEKAL plasma emission
code (Kaastra \& Mewe 1993; incorporating the Fe-L calculations of
Liedhal, Osterheld \& Goldstein 1995) and the photoelectric absorption
models of Balucinska-Church \& McCammon (1992). The ACISABS model was
used to account for time-dependent contamination along the instrument
light path.  Only data in the $0.8-7.0$ keV energy range were used for
the analysis.  The spectra for all annuli were modelled simultaneously
in order to determine the deprojected X-ray gas temperature profile
under the assumption of spherical symmetry. These data were then
combined with the observed surface brightness profile to constrain the
gas density profile.
 
We have fitted the electron density ($n_{e}$) profile for the central 
300 kpc of the cluster using a modified King model \cite{caval},

$$
n_{e}(r)=n_{0}(1+r^{2}/r_{c}^{2})^{-3\beta/2}. 
\eqno(1)
$$

\begin{figure}
\epsfig{file=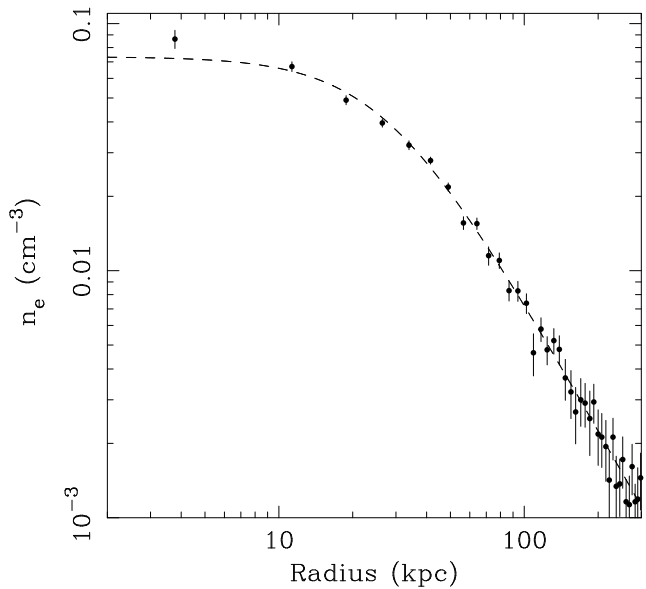, angle=-90, width=\linewidth}
\caption{Deprojected electron density profile of A\,2597 determined
from Chandra observations.  The dotted line shows the best-fitting
$\beta-$model.}
\label{betadensity}
\end{figure}

\noindent From this analysis we measure a central density, $n_{0}=7.35
\pm 0.32 \times 10^{-2}$ cm$^{-3}$, a core radius, $r_{c}=27.7 \pm
1.7$ kpc, and a slope parameter, $\beta=0.585 \pm 0.018$.  The model
fit, which gives an acceptable reduced $\chi^{2}$ value of 0.96, is
shown as a dotted line in Figure~\ref{betadensity}.  The modified King
model does not parameterize the gas density at large radii for the
A\,2597 cluster.  In this paper we only use the results of this model in
the central 200 kpc where this parametrization adequately describes
the gas density.

% Figure 1 -- Electron Density and Beta Model of A\,2597

\section{Results}\label{Results}

% Figure 2a,b,c  -- Total Intensity, C, X, U Band.
\begin{figure}
\vspace{20cm}
\includegraphics{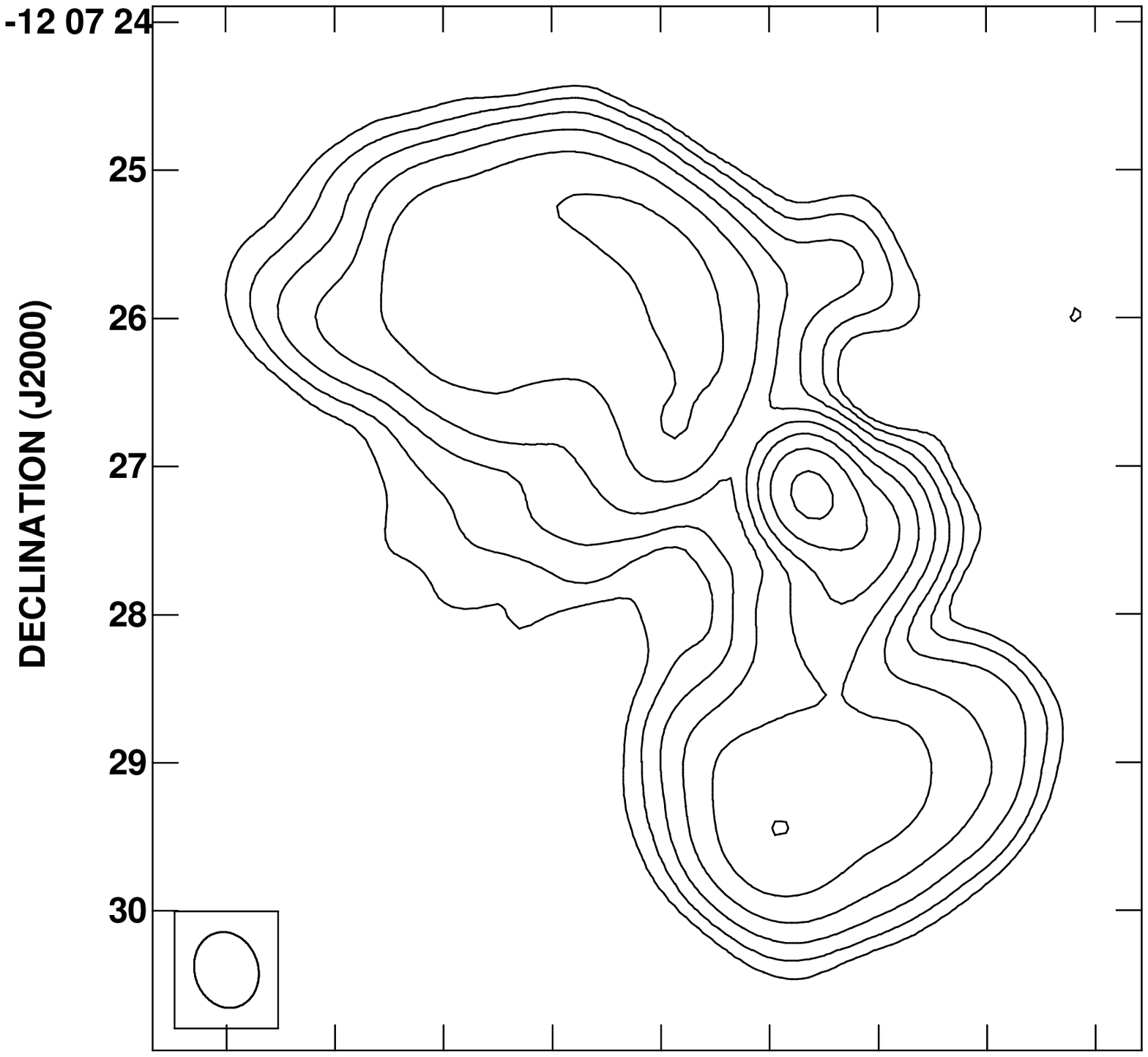}
\includegraphics{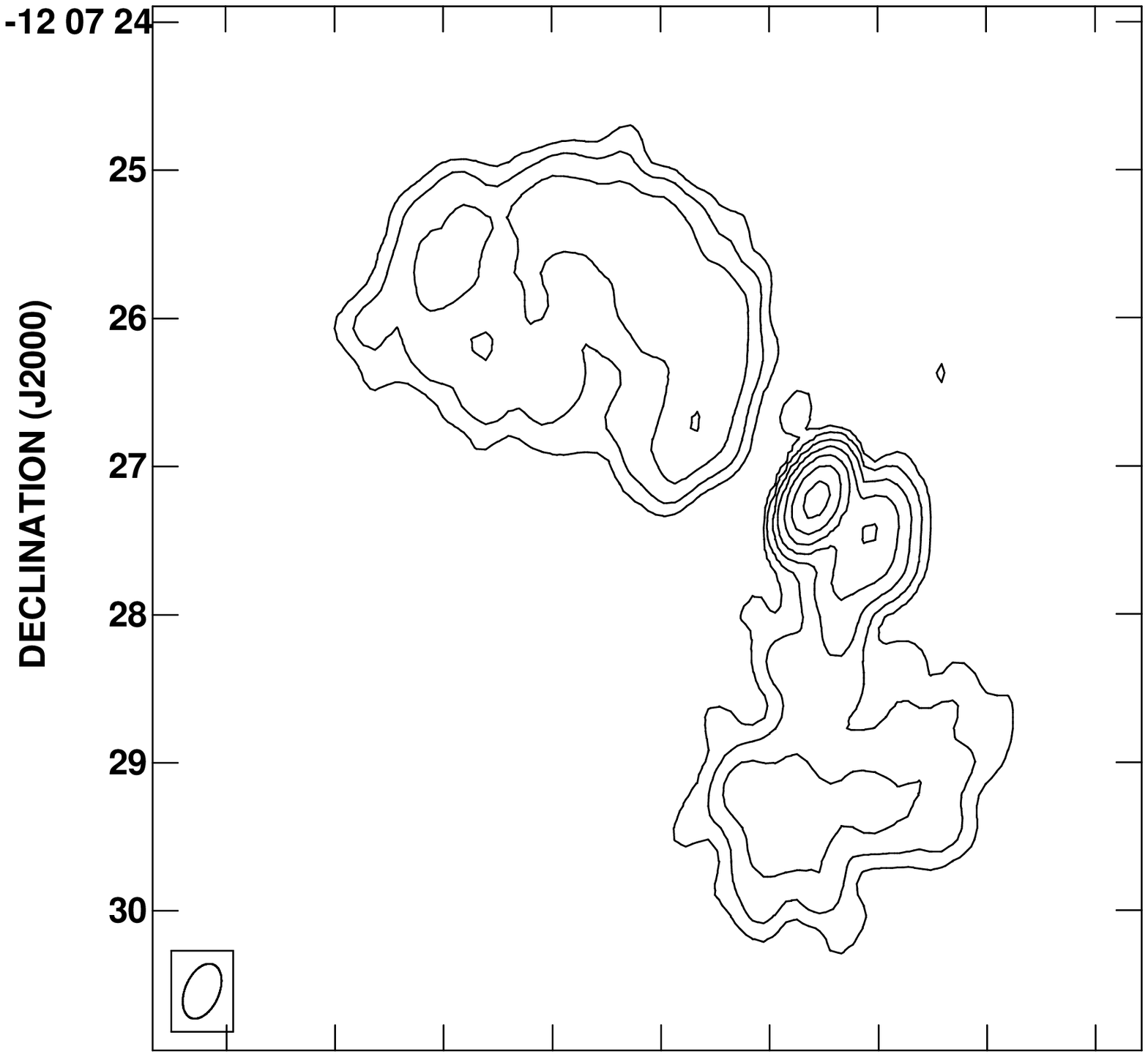}
\includegraphics{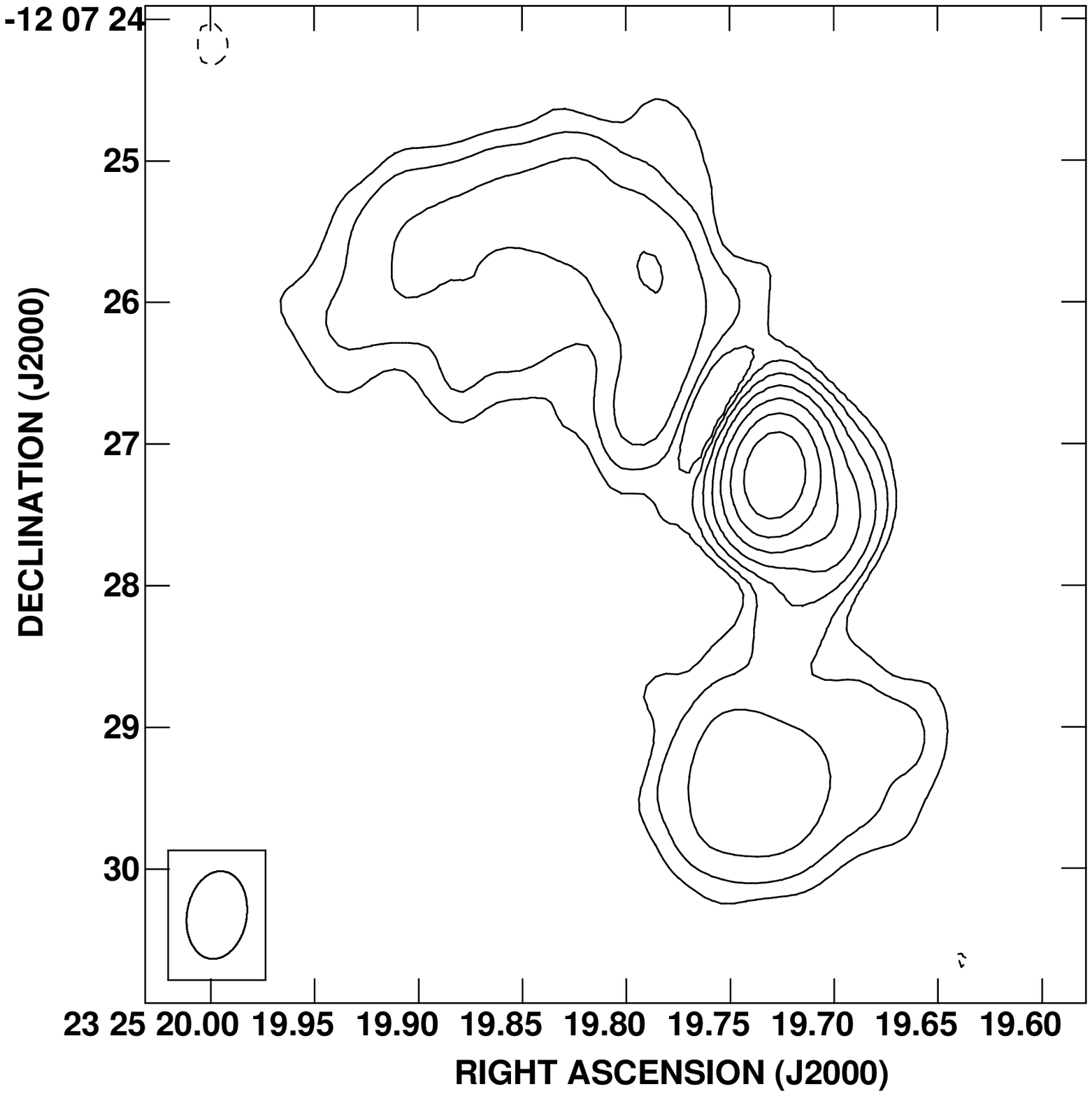}
\caption{From top to bottom, total intensity at 4985, 8377, and 14985
MHz.  Solid contours start at 0.35 mJy beam$^{-1}$ and increase by
factors of 2.  The peak contour flux is 59.3, 30.2, and 42.5 mJy
beam$^{-1}$ for the 4985, 8377, and 14985 MHz images,
respectively. The restoring beam is a Gaussian with FWHM 0.52 $\times$
0.43 arcsec$^2$ in P.A.  17\degr; 0.39 $\times$ 0.23 arcsec$^2$ in
P.A. $-$23\degr; and 0.62 $\times$ 0.42 arcsec$^2$ in P.A. $-$9\degr.
The respective noise in the images is 44, 91, and 86 $\mu$Jy
beam$^{-1}$.
\label{ctot}}
\end{figure}

Figure~\ref{ctot} shows total intensity contour images of
PKS\,2322$-$123 at 4985, 8377, and 14985 MHz.  In all three frequency
bands the source has an asymmetric structure about the core; there is
one area of diffuse emission located northeast of the core and another
area located directly south of the core.  We call these areas northern
and southern lobes.  Both the northern and southern lobes extend out
from the core to a distance of approximately 3.5\arcsec, or 5.4 kpc,
giving the source a total size of 7\arcsec, or 10.8 kpc.  There is a
component 0.5\arcsec\ southwest of the core that we refer to as the
inner jet.  The inner jet is most pronounced in the highest resolution
8 GHz image.  The 15 GHz image, because taken in the VLA's B
configuration, does not resolve the inner jet as well despite the fact
that it is our highest frequency.

% Figure 3 -- Spectral Index, XU over U
\begin{figure}
\epsfig{file=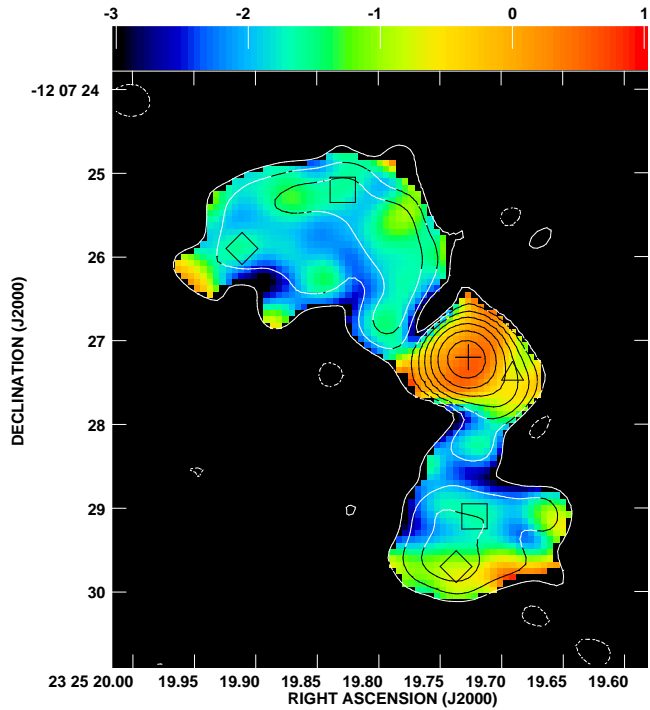, width=\linewidth}
\caption{Spectral index map made from 8377 MHz and 14985 MHz.  The
colorbar ranges from $\alpha^{14985}_{8377}=-3$ to $+$1.  Overlaid
contours show total intensity at 14985 MHz.  Solid contours start at
0.4 mJy beam$^{-1}$ and increase by factors of 2. Dashed contours
represent $-$0.4 mJy beam$^{-1}$.  The peak contour flux is 43.1 mJy
beam$^{-1}$. The restoring beam is a circular Gaussian with a FWHM
of 0.5\arcsec.  The overlaid symbols (square, diamond, triangle, and
cross) mark the locations used to show spectral index roll-off in
Figure~\ref{spectralindex}.}
\label{xuoveru}
\end{figure}

% Figure 4 -- Spectral Index, CU over U
\begin{figure}
\epsfig{file=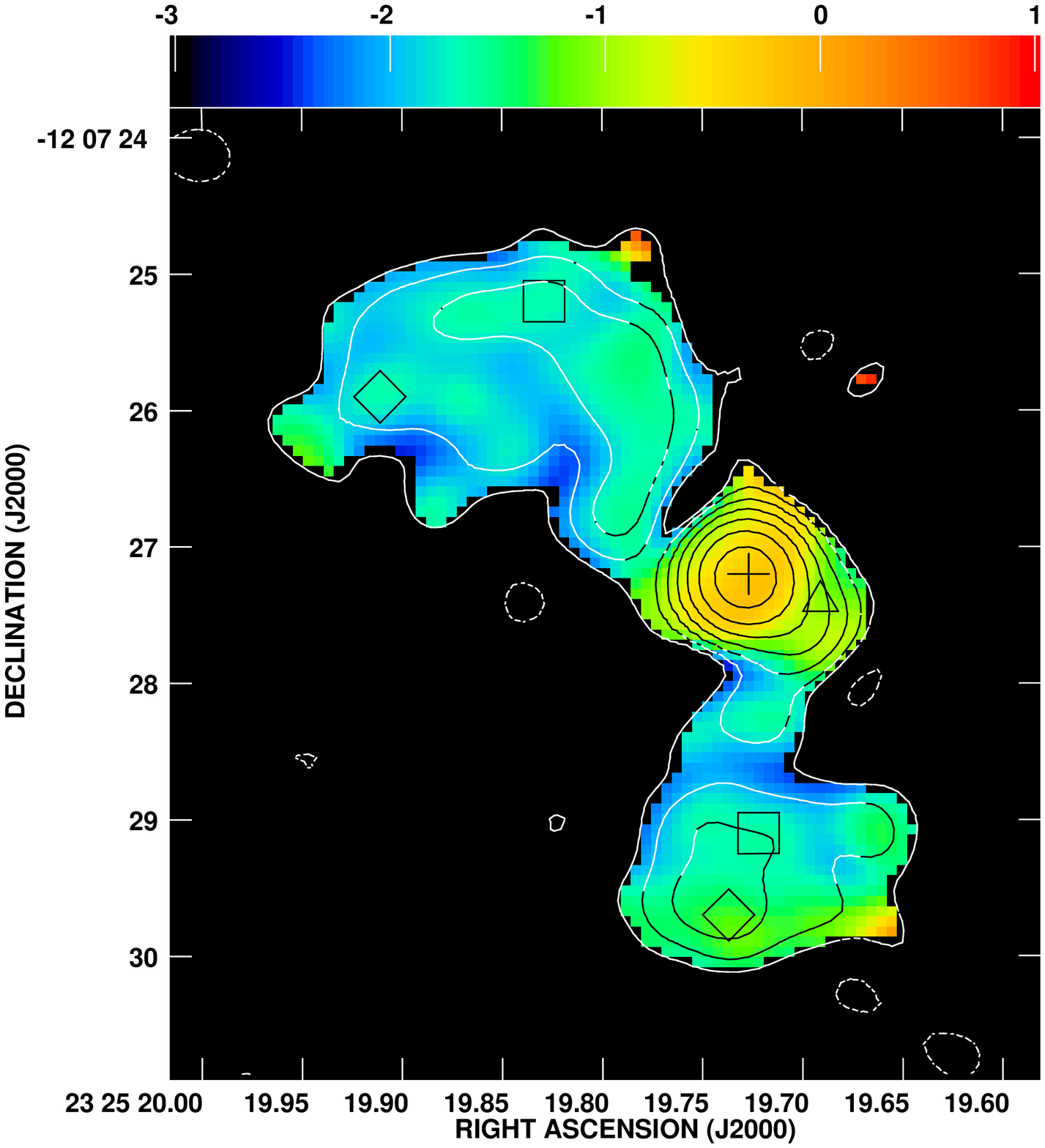, width=\linewidth}
\caption{Spectral index map made from 4985 MHz and 14985 MHz.  The
colorbar ranges from $\alpha^{14985}_{4985}=-3$ to $+$1.  Overlaid
contours show total intensity at 14985 MHz.  Solid contours start at
0.4 mJy beam$^{-1}$ and increase by factors of 2. Dashed contours
represent $-$0.4 mJy beam$^{-1}$.  The peak contour flux is 43.1 mJy
beam$^{-1}$. The restoring beam is a circular Gaussian with a FWHM
of 0.5\arcsec.  The overlaid symbols (square, diamond, triangle, and
cross) mark the locations used to show spectral index roll-off in
Figure~\ref{spectralindex}.}
\label{cuoveru}
\end{figure}

% Figure 5 -- Spectral Index Roll-Off
\begin{figure*}
\epsfig{file=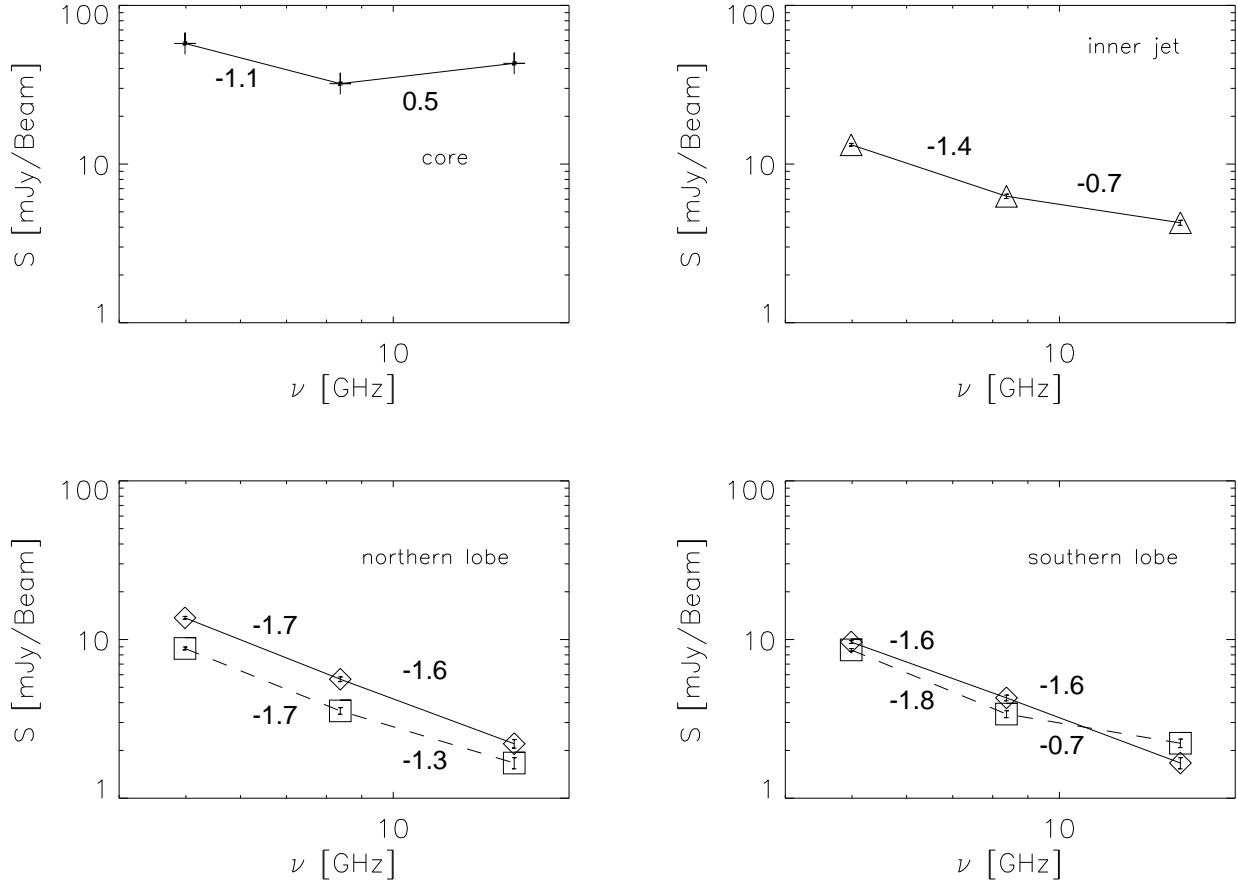, width=\linewidth}
\caption{Log-log plot of flux density versus frequency at 4985, 8377,
and 14985 MHz for six different source locations.  (8377 MHz
represents an average of the 6 X band frequencies used to calculate
the spectral index, and 14985 MHz is an average of the 2 U band
frequencies.)  The slope of the segmented line gives two spectral
indices, $\alpha^{8377}_{4985}$ and $\alpha^{14985}_{8377}$.  The
values of $\alpha$ are shown next to each segmented line.  The symbols
used (square, diamond, triangle, and cross) correspond to the symbols
plotted in Figure~\ref{xuoveru} and~\ref{cuoveru} which mark the
position on the source used in this spectral index analysis.  The
error bars shown are much smaller than plotted symbols, and represent
the RMS noise in the flux density maps and the 2\% accuracy of the
VLA.}
\label{spectralindex}
\end{figure*}

Figures~\ref{xuoveru} and ~\ref{cuoveru} show matched resolution
spectral index maps of PKS\,2322$-$123 in which
$\alpha^{14985}_{8377}$ and $\alpha^{14985}_{4985}$ are measured,
respectively.  Both maps are overlayed with total intensity contours
from the 14985 MHz image, and the color bar ranges from $\alpha=-3$ to
+1.  The northern and southern lobes have very steep spectra with
typical values of $\alpha=-1.8$, which is slightly steeper than the
spectral index of $-$1.5 previously observed by Sarazin \etal (1995).
The core and inner jet both have flatter spectra with typical values
in the core ranging from $\alpha=-1.2$ to $-$0.2 depending on the
frequencies.  It is interesting to note in Figure~\ref{xuoveru} that
the spectrum flattens in the southern region of the southern lobe.
The flattening is visible, but less pronounced in
Figure~\ref{cuoveru}.  No significant flattening of the spectrum
occurs in the northern lobe.  The cross and triangle symbols plotted
over the spectral index maps mark the total intensity centroid
positions of the core and inner jet.  The square and diamond symbols
mark four positions chosen for comparison in the northern and southern
lobes.  These symbols are used in Figure~\ref{spectralindex} to show
the difference in flux densities at our three observing bands; in each
of the four plots the spectral index is given by the slope of the
segmented line.  There appears to be no roll-off in the spectral index
at low or high frequencies.  Note that our 5 GHz, 8 GHz, and 15 GHz
observations were taken in 1991, 1994, and 1996, respectively, so that
the apparent dip in the spectral index plot of the compact core
component may be due to source variability on a time scale of 3 years.

\begin{figure*}
\epsfig{file=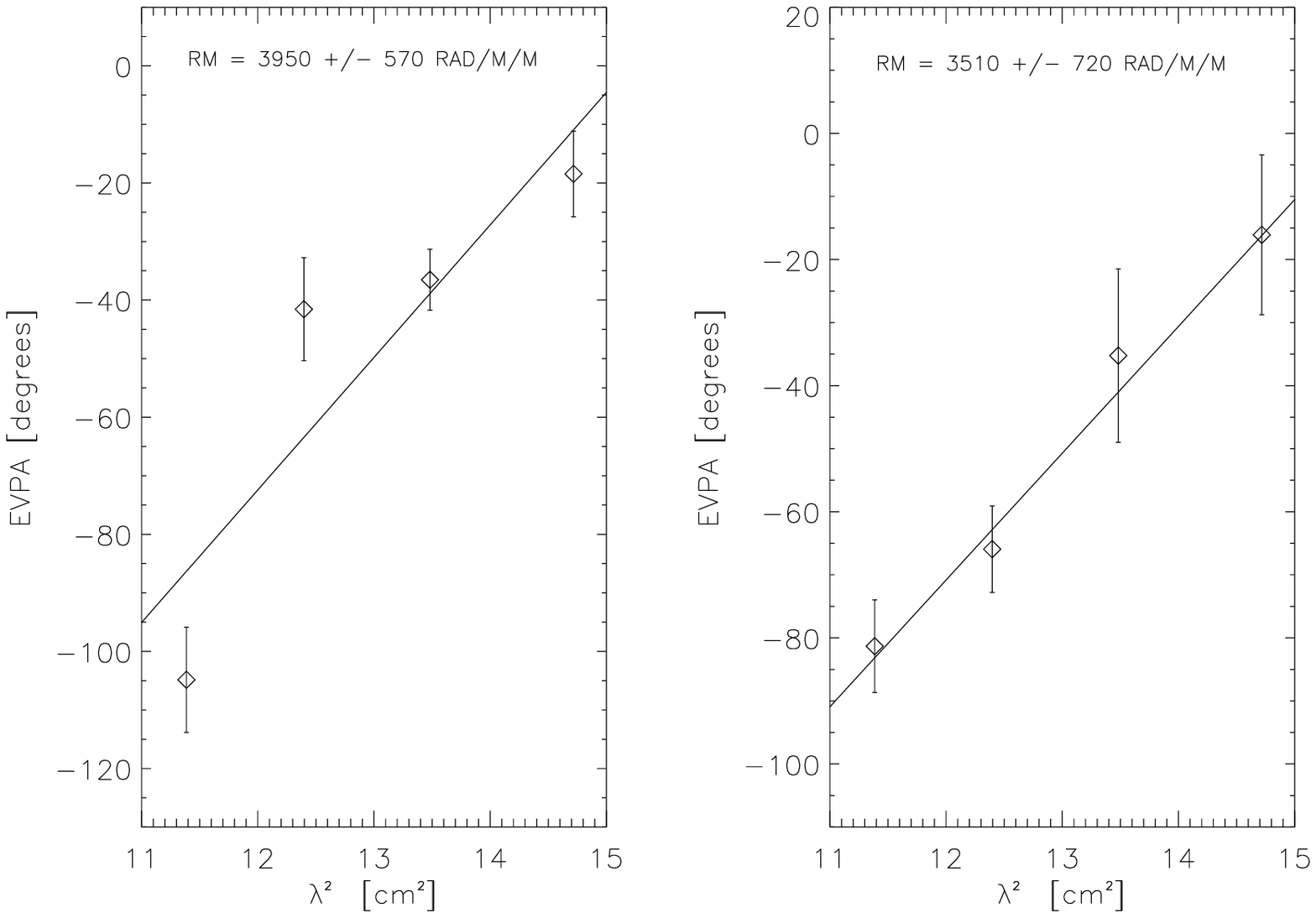, width=\linewidth}
\caption{Electric vector polarization angle versus square of
wavelength at 4 frequencies in the 8 GHz observing band for two source
locations. (left: RA $23^{\rmn{h}} 25^{\rmn{m}} 19\fs742$, Dec
$-12\degr 07\arcmin 29\farcs27$; right: RA $23^{\rmn{h}} 25^{\rmn{m}}
19\fs661$, Dec $-12\degr 07\arcmin 28\farcs471$) Error bars show
1$\sigma$ taken from the $\chi$ noise maps.  Solid line is a
least-squares-fit to the data.  The slope of the solid line gives the
rotation measure.  For the eastern location, RM=$3950 \pm 570$ \radm;
for the western location, RM=$3510 \pm 720$ \radm.}
\label{rmfit}
\end{figure*}

Figure~\ref{rmoverlay} shows the rotation measure distribution created
from 7815, 8165, 8515, and 8885 MHz.  The color bar ranges from RM
values of 800 \radm\ to 6800 \radm.  Pixels with an error in the
$\chi$ map of greater than 20\degr for any of the four frequencies have
been flagged, and pixels corresponding to source locations with
less than 200 $\mu$Jy beam$^{-1}$ of total intensity flux at 8165 MHz
are not shown.  The overlayed contour lines show total intensity at
8885 MHz.  Note that the polarized flux only appears in a small region
of the southern lobe.  Figure~\ref{rmfit} shows RM fits for two pixels
located in the cluster of pixels in the southern lobe.  The
fits result in rotation measures of $3950 \pm 570$ \radm\ and $3510
\pm 720$ \radm, which are typical values for this source.  Using the
62 pixels clustered together in the southern lobe, and excluding four
pixels due to bad RM fits, we produce a histogram of the rotation
measure in Figure~\ref{rmhistogram}.  The histogram shows a mean RM
value of 3620 \radm\ and dispersion of 1080 \radm.  Because we have
only 62 unflagged pixels in our RM map spanning about two beam widths,
we consider there to be only one coherent area of the source with an
RM value, namely the mean value.  Although an RM dispersion of
$\sim$1000 \radm\ is not atypical, due to the small number of
clustered pixels in our histogram we do not consider this dispersion
to be representative of the physical spread in RMs across the source.
Instead, this dispersion most likely results from our low signal and
is due to errors in the RM determination which are estimated to be $\sim$700
\radm.

\section{Discussion}\label{discussion}

\subsection{Magnetic Field Strengths and Topologies}\label{Bfield}

At its high galactic latitude ($b=-64.85$\degr), A\,2597 is far from the
plane of our galaxy, and at its location we expect galactic RMs of
less than 30 \radm\ \cite{simard}.  Therefore our observed RMs located
at a projected distance of 2\arcsec\ from the cluster center are most
likely the result of Faraday rotation from a magnetized cluster gas,
or Faraday rotation from a magnetic field within the cD galaxy which
is located at the center of the cluster.  It is also possible that the
RMs are the result of interaction between the radio galaxy and the
surrounding ICM \cite{bicknell}, however with polarized flux over a
region of only 5 kpc, it is difficult to determine which of these
scenarios produces the dominant field.  Based on similar findings of
high RMs in cooling core clusters with embedded sources of very
different morphologies \cite{carilli}, we favor a cluster magnetic
field interpretation.

\subsubsection{A Uniform Cluster Magnetic Field}

Assuming that the rotation measure is the result of a magnetized cluster
gas, rotation measures can be related to the line-of-sight magnetic
field, $B_{\|}$, by

$$ RM = 812\int\limits_0^L n_{\rm e} B_{\|} {\rm d}l ~{\rm
  radians~m}^{-2}~,
\eqno(2)
$$

\noindent where $B_{\|}$ is measured in $\mu$G, $n_{e}$ in cm$^{-3}$,
d$l$ in kpc, and the upper limit of integration, $L$, is the distance
from the emitting source to the end of the path through the Faraday
screen along the line of sight.

We can estimate the minimum magnetic field needed to produce our
observed RMs by making the simplest assumptions of a constant electron
density, $n_{0}$, and a uniform magnetic field out to a distance of
the core radius, $r_{c}$.  Using the values calculated in
\S\ref{XrayObs}, and setting the rotation measure equal to our mean
observed RM of 3620 \radm, we find $B_{\|}=2.1 \mu$G.  This is a lower
limit on the magnetic field strength because any field reversal along
the line of sight would require a larger magnetic field to produce the
same observed RM.  Note that for $r>r_{c}$ the density falls off as
$r^{-2}$ and, assuming flux conservation, the magnetic field scales
with $n^{2/3}$.  This results in the rotation measure scaling with
$r^{-7/3}$ so that the contribution to the RM is dominated by the
cluster center.

\subsubsection{A Tangled Cluster Magnetic Field}

% Figure 6 -- Rotation Measure image of PKS\,2322$-$123
\begin{figure}
\epsfig{file=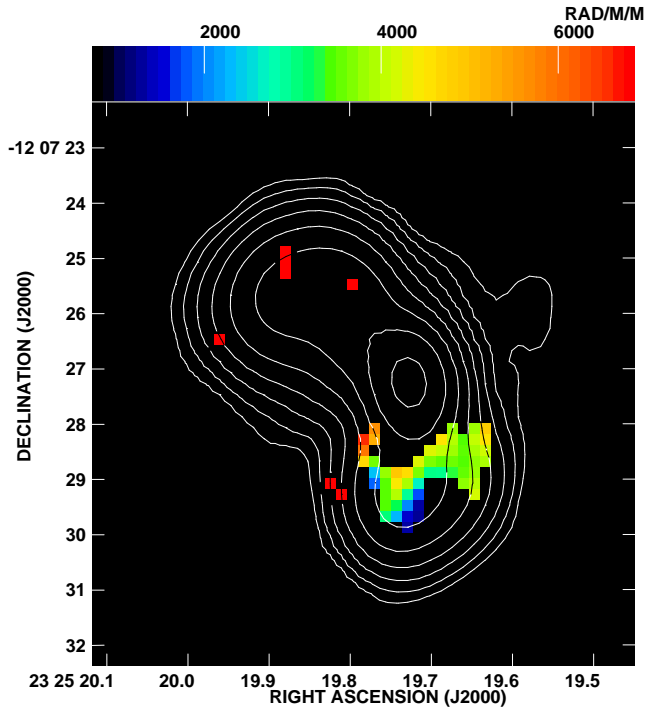, width=\linewidth}
\caption{Rotation measure image of PKS\,2322$-$123 with contours of
total intensity at 8885 MHz overlaid.  The colorbar ranges from 800 to
6800 \radm.  Rotation measure fits were performed using 7815, 8165,
8515, and 8885 MHz data.  Pixels corresponding to a total intensity of
less than 200 $\mu$Jy beam$^{-1}$ (%5 \sigma$) at 8165 MHz are not
shown.  Contours start at 0.35 mJy beam$^{-1}$ and increase by factors
of 2.  The peak contour flux is 56.2 mJy beam$^{-1}$.  The restoring
beam is a Gaussian with a semimajor axis of 1.5\arcsec, a semiminor
axis of 1.0\arcsec, and a position angle of 0\degr.}
\label{rmoverlay}
\end{figure}

% Figure 7 -- Rotation Measure Fit example for 2 points

A more realistic estimate of the magnetic field assumes random field
reversals on the order of a coherence length.  Although in
figure~\ref{rmoverlay} the RM values are purely positive so that field
reversals are not apparent, from the patchiness of the RM
distributions observed in other sources, magnetic fields ordered on
scales of $\sim$5 kpc are common \cite{carilli}.  Due to the small
region of polarized intensity in PKS\,2322$-$123 it is difficult to
determine a suitable cell size by estimating the coherence length in
the RM map.  Therefore we use a cell size of 5 kpc, a typical value
for both cooling core and noncooling core clusters \cite{carilli}.
Using the relation derived by Felten (1996), we can estimate the
magnetic field assuming our calculated density distribution that
follows a $\beta-$profile, and a magnetic field topology with cells of
constant size and magnetic field strength, but random orientation.

$$
\sigma_{\rm RM} = {{{\rm K} B~n_0~r_c^{1/2} l^{1/2}}
\over{(1 + r^2/r_c^2)^{(6\beta - 1)/4}}}
\sqrt{{\Gamma(3\beta - 0.5)}\over{{\Gamma(3\beta)}}}
\eqno(3)
$$

\noindent In this equation, $l$ is the cell size in kpc, $r$ is the
distance (in kpc) of the radio source from the cluster center,
$\Gamma$ is the Gamma function, $B$ is given in $\mu$G and is related
to the component along the line of sight by $B=\sqrt{3}B_{\|}$, and K
is a factor that depends on the location of the radio source along the
line of sight through the cluster: K=624 rad $\mu$G$^{-1}$ if the
source is beyond the cluster and K=441 rad $\mu$G$^{-1}$ if the source
is halfway through the cluster.  The parameters $r_{c}$, $n_{0}$, and
$\beta$ are as described in \S\ref{XrayObs}.  Using values given in
\S\ref{XrayObs} and in Table~\ref{SourceProperties}, and a cell size
of 5 kpc, we estimate the cluster magnetic field to be 2.9 $\mu$G.
This estimate which makes use of a tangled magnetic field model gives
a line-of-sight magnetic field of $B_{\|}=1.7 \mu$G, in agreement with
the previous field estimate which assumed a uniform field topology.
The fact that these two magnetic field topologies yield similar
line-of-sight magnetic field strengths does not suggest any preferred
field topology.  Rather, it is more likely that $\sigma_{\rm RM}$ used
in Equation 3 above is not a robust measurement of the true RM
dispersion since the region covered is only a few beamwidths and the
SNR is low.  The dispersion is probably more indicative of the quality
of the RM fit.  Therefore, we conclude only that the cluster magnetic
field has a minimum strength of 2.1 $\mu$G.  We note that this
magnetic field strength is somewhat lower than the 5 to 10 $\mu$G
typically found in cooling-core clusters \cite{carilli} which leads us
to believe that a more complicated field topology is present, or that
the RM dispersion has been underestimated.

\subsection{The Nature of the Radio Source}

The radio source PKS\,2322$-$123 is located at the center of the
galaxy cluster A\,2597, and is coincident with a cD galaxy.  Its small
physical size, steep-spectrum lobes, and disturbed jet morphology all
suggest the radio emission is confined by the ambient gas.  This is
similar to what is seen in the cluster center radio sources
PKS~1246-410 in the Centaurus cluster (Taylor, Fabian, \& Allen 2002), and
PKS~0745-191 in the cluster of the same name \cite{tay94}.  The steep
spectral indices of both the northern and southern lobes can be
explained by a trapping of electrons such that low energy particles
dominate the emission \cite{mye85}.  In the past many people have
classified A\,2597 as a cooling-core cluster.  Our analysis confirms
the presence of a high central gas density in the cluster
(Figure~\ref{betadensity}) and a short central cooling time; $t_{\rm
cool} < 10^9$ yr within $r\sim 30$ kpc of the central radio source.

\begin{figure}
\epsfig{file=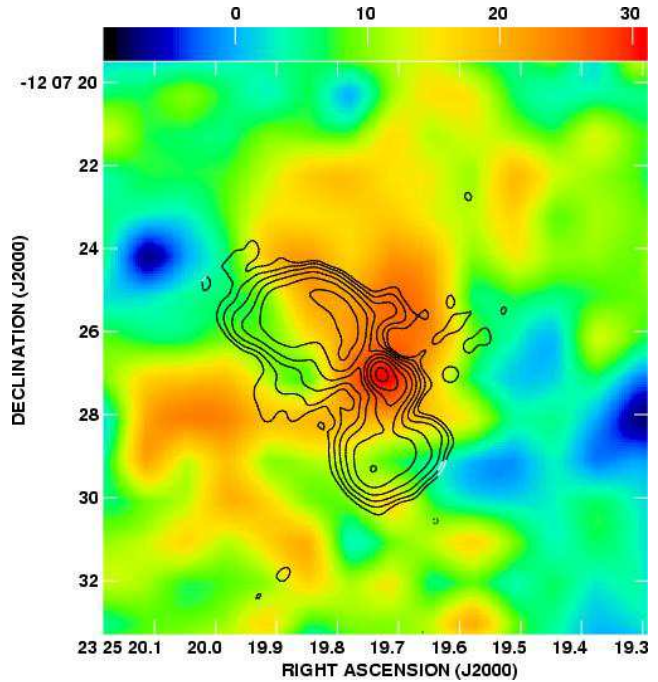, width=\linewidth}
\caption{Unsharp masked image of the X-ray emission from the
inner part of the A\,2597 cluster at 1.1 arcsecond resolution, overlaid
on the 5 GHz VLA image.}
\label{unsharp}
\end{figure}

The flattening of the spectral index in the southern part of the
southern lobe may be caused by interactions between the radio jet and
the surrounding ICM.  As the jet moves south hitting the X-ray gas,
some of the jet's kinetic energy may reaccelerate the relativistic
particles at the boundary layer, causing the spectrum to flatten.  A
similar situation may be responsible for the flatter spectrum of the
northern radio lobe along its edge.  To investigate this idea further
we have constructed a 1 arcsec$^2$ pixel image from the X-ray data
(reprocessed using relaxed screening criteria that exclude only the
largest background flares, resulting in a net exposure time of 28ks).
We first lightly smoothed this image, convolving it with a 1.2
arcsecond FWHM Gaussian kernel. This lightly smoothed image was then
subtracted from a more heavily smoothed image, constructed by
convolving the raw image with a 12 FWHM arcsecond Gaussian kernel. The
resulting unsharp-masked image is shown in Figure~\ref{unsharp}.  The
X-ray and radio peaks are nicely aligned without any correction to the
X-ray position beyond the standard calibration.  Enhanced X-ray
emission is seen both along the northern edge of the northern lobe,
and at the tip of the southern lobe -- both regions where the spectral
index is seen to flatten.  In addition to this evidence of interaction
between the radio jet and the ICM at the edges of the lobes, Koekemoer
et al. (1999) find H$\alpha$ emission-line filaments that trace the
northern and southern lobes on the northwest and southeast edges,
respectively.  These edge-brightened arcs further suggest physical
interactions at the boundary layer.

% Figure 8 -- Rotation Measure distribution of PKS\,2322$-$123 vers. 1
\begin{figure}
\epsfig{file=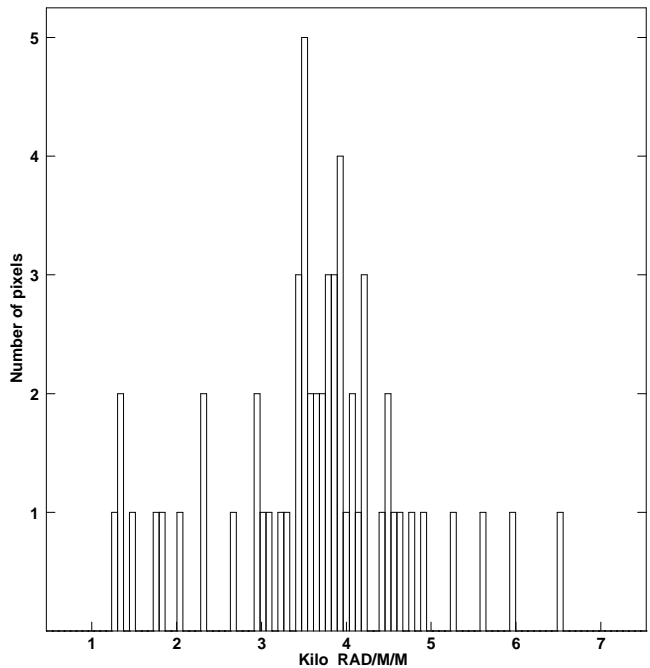, width=\linewidth}
\caption{Rotation measure distribution for the 58 pixels clustered
south of the core in Figure~\ref{rmoverlay}.  (Four pixels have been
discarded due to bad rotation measure fits.)  The mean value is
3620\radm\ with dispersion of 1080 \radm.}
\label{rmhistogram}
\end{figure}

% Figure 9 -- Unsharp masked image of the X-ray overlaid with radio

One explanation for the lack of detectable polarization in the northern
lobe is that we have a greater Faraday depth towards this lobe,
resulting in beamwidth depolarization.  We calculate that at 8 GHz a
Faraday rotation gradient of $\sim$350 \radm\ kpc$^{-1}$ is sufficient
to rotate $\chi$ by 1 radian over our 2 kpc beam, thus causing
depolarization over the beam width.  This would require RMs on the
order of 700 \radm, 5 times lower than the observed RMs in the
southern lobe.  Thus Faraday depolarization can easily explain the
lack of observed polarized flux in the northern lobe, and the low
amounts of polarized flux in the southern lobe as well.

\begin{figure}
\epsfig{file=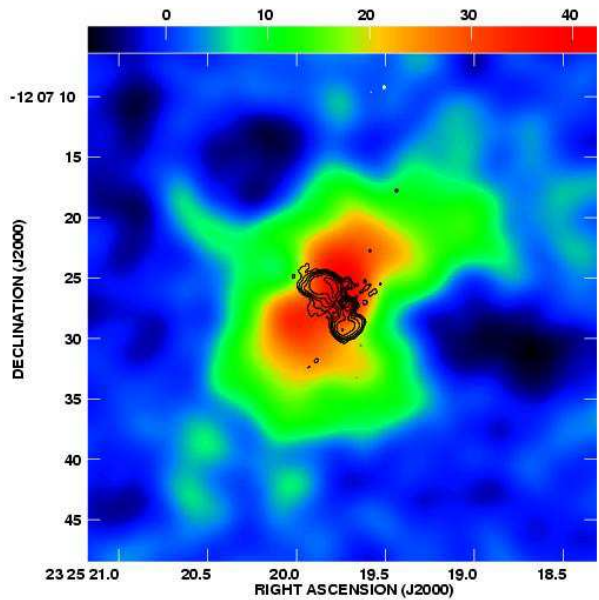, width=\linewidth}
\caption{Unsharp masked image of the X-ray emission from the inner
part of A\,2597 cluster at 3 arcsecond resolution, overlaid on the 5
GHz VLA image.  Notice the cavities to the northeast and southwest of
the radio core.}
\label{cavities}
\end{figure}

Two low density cavities were detected some 10s of arcseconds
northeast and southwest of the radio core by McNamara (2001).  These
two cavities can be clearly seen in Figure~\ref{cavities}.  According
to McNamara (2001) the low density cavities are at a projected
distance of roughly 30 kpc from the center of the cluster.  The X-ray
cavities detected by McNamara \etal (2001) suggest some interesting
jet movement.  McNamara (2001) gives evidence supporting the idea that
the cavities are associated with a radio outburst that occurred 50 --
100 Myr ago.  The recent discovery of 330 MHz radio emission in the
western cavity by Clarke et al. (in prep) confirms the relic radio
outburst theory.  We suggest that the same radio jet that made the
southern lobe in our current observations was oriented southwest of
the core in that previous radio episode and was deflected to the south
at a later time.  It is possible that as the relic radio bubble rose
outward, the X-ray gas filling in the volume at low radii provided the
additional pressure that caused the jet's deflection.  VLBA
observations of PKS\,2322$-$123 show that on milliarcsecond scales the
radio source is symmetric and has a northeast, southwest orientation
\cite{tay99}, providing further evidence that the southern jet begins
in a southwesterly direction and is deflected to the south.  Our
observations of a bright inner jet southwest of the core (most easily
seen in Figure~\ref{ctot}b) are consistent with the VLBA observations
and the theory of a deflected jet.  The inner jet's relatively high
flux density may be caused by interactions at the boundary layer at
the deflection point.  If the southern lobe of the radio source is
propagating toward us, then we note that the observed angle through
which the southern jet is deflected may be larger than the actual
deflection angle due to projection effects.  Finally, based on the
detection of \HI\ absorption toward the core and northeastern inner
jet by Taylor \etal(1999), we suggest a source morphology in which the
southern, deflected jet is on the approaching side.  The fact that
\HI\ was not detected toward the southwestern jet can be explained if
the \HI\ is in a torus around the core, and the southwestern jet is on
the approaching side.

% Figure 10 -- Large X-ray/radio overlay to show cavities mentioned in McNamara.

If our deflected jet scenario is correct, then we may expect to see a
pressure imbalance between the thermal X-ray gas and non-thermal gas
in the radio lobes.  We might expect the X-ray gas to have equal or
greater pressure than the non-thermal gas; this would be consistent
with our analysis of a small, confined radio jet.  But in fact, using
a minimum energy condition described by Miley (1980), we calculate the
minimum pressure of the radio source to be $\sim1.4 \times 10^{-9}$
erg cm$^{-3}$ in both the northern and southern lobes, and $\sim7.0
\times 10^{-9}$ erg cm$^{-3}$ in the inner jet.  The lobe values are
in rough agreement with our calculated X-ray gas pressure of $\sim7
\times 10^{-10}$ erg cm$^{-3}$ in the northern and southern
lobes. However, the minimum pressure in the inner radio jet is
approximately an order of magnitude larger than the X-ray gas pressure
at that radius ($\sim8 \times 10^{-10}$ erg cm$^{-3}$ at $r \sim 1$
kpc).  If our assumption of the minimum energy condition is incorrect,
and equipartition cannot be assumed, then the radio lobes would have a
higher pressure than that which we calculated.  Dunn and Fabian (2004)
find that though the assumption of equipartition is sometimes correct,
often it fails for reasons not well understood.

An alternative theory to that of the deflected southern jet is a
precessing radio jet as described by Gower \etal (1982).  If
precession rather than deflection is the cause of the misalignment
between the X-ray cavity and southern radio lobe, then the inner jet
we observe with the VLA and the symmetry observed with the VLBA are
evidence that the jet has now returned to the orientation it was in 50
-- 100 Myr ago.

\section{Conclusions}

We detect polarized emission from the southern lobe of the radio
source PKS\,2322$-$123, which is embedded in the cooling-core cluster
A\,2597.  We calculate this region of polarized flux to have a Faraday
rotation measure of 3620 \radm, which is consistent with the high RMs
observed in other embedded radio sources \cite{tay94,carilli}.  Assuming both
uniform and tangled magnetic field topologies, we use our calculated
RMs to estimate a minimum cluster magnetic field of $2.1 \ \mu$G,
which is low compared to the 5 -- $10 \ \mu$G typically found in
galaxy clusters using similar methods \cite{carilli}.  Because
polarized flux was only detected over a very small ($\sim3$ kpc)
region, we believe that our low cluster magnetic field estimate is due
to an underestimate in the RM dispersion.  Future, more sensitive
observations made with the Expanded VLA should reveal more polarized
flux, allowing us to better estimate the RM dispersion.

We show two spectral index maps in which we measure steep spectrum
lobes with $\alpha=-1.8$ and a flatter spectrum core with $\alpha$
ranging from $-$1.2 to $-$0.2 depending on the frequencies. We suggest
that the flattening could be due to a boundary layer interaction with
the ISM of the host galaxy.  Comparison with the Chandra X-ray
observations on the arcsecond scale show some weak evidence for
density enhancements at the positions where the radio spectral index
flattens. Future low frequency radio observations are needed to test
this interaction hypothesis.  Low frequency observations will also be
interesting to determine the spectral curvature; we see no roll-off in
the spectral index at 5, 8, and 15 GHz.

Based on our comparison of our radio observations with previous X-ray
\cite{mcnam} and VLBA observations \cite{tay99}, we advocate a source
orientation and morphology in which the southern jet has been
deflected south from its original southwestern orientation, and is
on the approaching side.  Future proper motion studies on
milli-arcsecond scales could help determine whether this proposed
source orientation is correct.  If the western VLBI jet is propagating
toward us relativistically, then we would expect to see faster motions
on that side due to relativistic effects.

\section*{Acknowledgments}
We thank Tracy Clarke and an anonymous referee for constructive
comments.  SWA thanks the Royal Society for support.  Support for this
work was provided by the National Aeronautics and Space Administration
through Chandra Award Number GO4-5134X issued by the Chandra X-ray
Observatory Center, which is operated by the Smithsonian Astrophysical
Observatory for and on behalf of the National Aeronautics Space
Administration under contract NAS8-03060.  This research has made use
of the NASA/IPAC Extragalactic Database (NED) which is operated by the
Jet Propulsion Laboratory, Caltech, under contract with NASA.  The
National Radio Astronomy Observatory is a facility of the National
Science Foundation operated under a cooperative agreement by
Associated Universities, Inc.

\bsp
\label{lastpage}
\end{document}